\documentclass[pre,aps,showpacs]{revtex4}
\usepackage{epsfig}
\newcommand{\be}{\begin{equation}}
\newcommand{\ee}{\end{equation}}
\newcommand{\bn}{\begin{eqnarray}}
\newcommand{\en}{\end{eqnarray}}
\newcommand{\bd}{\begin{displaymath}}
\newcommand{\ed}{\end{displaymath}}
\newcommand{\bnn}{\begin{eqnarray*}}
\newcommand{\enn}{\end{eqnarray*}}
\begin{document}
\title{The simulation of the activity dependent neural network growth}
\author{F. Gafarov$^1$\footnote{e-mail: fgafarov@yandex.ru}, N.
Khusnutdinov$^{1,2}$\footnote{e-mail: knr@kazan-spu.ru} and F.
Galimyanov$^1$}
\address{1) Department of Physics, Tatar State  University of Humanity and
Pedagogic, 420021, Kazan, Tatarstan 2, Russia \\
2) Department of Physics, Kazan State University, Kremlevskaya 18, 420008,
Kazan, Russia}
%\maketitle
% \begin{history}
% \received{7 September 2008}
% \revised{}
% \end{history}

\begin{abstract}
It is currently accepted that cortical maps are dynamic
constructions that are altered in response to external input.
Experience-dependent structural changes in cortical microcurcuts
lead to changes of activity, i.e. to changes in information encoded.
Specific patterns of external stimulation can lead to creation of
new synaptic connections between neurons. The calcium influxes
controlled by neuronal activity regulate the processes of
neurotrophic factors released by neurons, growth cones movement and
synapse differentiation in developing neural systems. We propose a
model for description and investigation of the activity dependent
development of neural networks. The dynamics of the network parameters
(activity, diffusion of axon guidance chemicals, growth cone
position) is described by a closed set of differential equations.
{The model presented here describes the development
of neural networks under the assumption of activity dependent axon
guidance molecules. Numerical simulation shows that morpholess neurons compromise the
development of cortical connectivity.}

\end{abstract}

\maketitle
\section{Introduction}
Neural networks are not constant structures. Modification in neural
nets leads to changes in mapping of input signal to output. The most
explored type of neural plasticity is synaptic plasticity.
Synaptic plasticity deals with modifications of connection
strength between neurons. It is an activity dependent process, and
synaptic efficacy modifications depend on activity of
postsynaptic and presynaptic neurons \cite{Hebb,Kempter}.
The second type of plasticity is known as structural plasticity.
Structural plasticity deals with anatomical structure of neurons,
and connections between neurons. The anatomical
structures of neurons are subject to variation, and new connections between
neurons can be established or deleted \cite{ChklovskiiMelSvoboda} in
the course of development of neuron net. Structural plasticity as
well as synaptic plasticity is a permanent and activity dependent
process. Because it is an activity dependent process it can be the
basis of learning. Activity-dependent modifications of neural
circuits lead to changes in activity pattern of whole network. The
geometrical properties of neurons should be considered in the
theoretical investigations of structural plasticity of neural netsworks.
The network must be considered as a system of neurons in
three dimensional neuropil where neurons
communicate with each other. The geometrical properties of neurons
and the neuropil where neurons are constituent elements of networks.
The neural networks generate activity patterns
which depend on the intrinsic state of individual neuron and
external influences on the system. This activity influences
structural plasticity process. An external signal changes the
activity pattern and leads to creation of new connections i.e. network
will be modified according to external information.

The neurons without connections can be considered a system of neurons in three
dimensional space. At the beginning, the neurons interact only by emission of
chemicals. Even no connections between neurons exist it may still be considered a
system. With time new connections between neurons appear and the system will possess' new
properties and neurons can make influence on the activity of each other neurons.

In vivo and in vitro neurons self-organize into networks \cite{Shefi,Segev2003}.
Mature networks can be considered a result
of activity-dependent dynamical wiring process. Single neurons have
properties that form the assembly into networks. Calcium plays
the most important role in wiring process
\cite{Garyantes,Neely,Henley}. Neuronal activity via
voltage-dependent calcium channels provide influx of calcium through
membrane. Intracellular calcium activates and regulates different
intercellular processes which influences growth cone movement,
axon elongation, neurotrophin release, synaptogenesis, among other molecular mechanisms. The
molecular mechanisms of most of these processes remains unclear
and are subjects of many experimental studies. Wiring process is
controlled by intrinsic neuronal activity and neural activity is
caused by sensory experience (externals signals). External signals regulate neuronal activity and leads formation of
the wiring between neurons. This adaptation leads to
different neural activity even under constant sensory input,
enabling the building of more complex representation and leading to
progressive cognitive development.

Three levels of neuronal response to external signals can be
considered: (i) induced spikes, (ii) synaptical plasticity and (iii)
structural plasticity. These processes have different time scales
namely, spikes -- milliseconds, synaptical plasticity -- hours,
structural plasticity  -- days. In the present paper we take into
account the first and third levels of consideration. In the future
we plan to include in our model the second level.

In this paper we present a mathematical model of the neural activity, underlying the development of neural networks.
The basis of our diffusion model is the experimental support of physiological and anatomical data.
Numerical simulations show the neural network growth, and how
neural activity controls this process.  The results may be used for
experimental verification of the neural network growth and for
conditions of new experiments. Some parameters of our
model have no experimental basis (e.g. dependence of amount of AGM released by
neurons on activity) and may needs to be verified in future experiments.

Models of axon guidance have been considered in detail by
Hentschel and van Ooyen \cite{Hentshel}. Three types of
diffusible molecules have been considered: a chemoattractant
released by target cells, a chemoattractant released by the axonal
growth cones and chemorepellant released by axonal growth cones. Two
cases were considered namely, (i) diffusible signals only, and (ii) contact
interactions with diffusible signals. It was shown
that target-derived chemoattractant controls axon guidance, the
axon-derived chemoattractant and chemorepellant control bundling and
debundling. The dynamics of chemicals concentrations are described by
standard diffusion equation. Every chemical has its own diffusion constant, the
release rate and degradation parameter. The release rate constants of chemicals
has no dependence on the state of the cell which releases them. In the
framework of the model the growth cones response to the concentration
gradients of chemicals and total response of the growth cone are the result of
two attractive and one repulsive concentration gradient.

Our model, in the part concerning AGM's diffusion and growth cone movement, is
based on the above mentioned models. For simplicity we consider only one type
of chemoattractant. For description of the movement of growth cone one uses more
complicated equations. The main difference is that, the release of
chemoattractant is controlled by activity of the target cells, and the growth
rate of the growth cones depends on the activity state of the neuron with
growing axon. Our model can be considered as generalization in the point of
neuron activity, of the model presented in the  \cite{Hentshel}.

 Some models of activity-dependent neural network development have already been
considered. First of all the model suggested by van Ooyen and
collaborators \cite{Ooyen1,Ooyen2}. The model consists of initially
disconnected neurons, modeled as neuritic field, which are organized
into a network under influence of their internal activity. The
growth of neurites are connected with Ca$^{2+}$ concentration inside
the cell. Therefore, the growth of neurites depends on their own
level of activity, and the neurons become connected when their
fields overlap. According to this model the high level of activity
causes neurites to retract, whereas low level allows further
outgrowth. From mathematical point of view they used a system of
coupled differential equations for neural activity and connection
strength. They showed, that spatial distribution of the cells can
create connectivity patterns in which hysteresis appears and complex
periodic behavior takes place \cite{Ooyen1,Ooyen2}.

The Segev with collaborators presented a model that
incorporates stationary units representing the cells soma and
communicating walkers representing the growth cones. The dynamics of
the walker's internal energy is controlled by the soma, and they
migrate in response to chemorepulsive and chemoattractive glues
emitted by the soma and communicate with each other and with the
soma by means of chemotactic feedback \cite{Segev2000}.

Our model is based on axon guidance by extracellular signals
released by other neurons. Our consideration is based on the
diffusion of the AGM. We already considered this approach before in
simple form \cite{Gafarov} with binary neurons and without detail
consideration of the process of diffusion. Also, the parameters of the
model have had no connection with reality and were taken from the
mathematical point of view to obtain suitable results. The main
novelty in our approach is that we consider the activity-dependence of the
processes underlaying neural network growth in more detail.

\section{Neurobiological motivation}
\subsection{Axon growth and guidance}

Axon growth requires the interplay of many processes: producing cytoplasmic and
membrane elements, shipping these building blocks to the right compartment and
inserting them into the growing axon and coordination
of all these processes.

The tips of growing axons equipped by a very specialized structure, called
growth cone, which is specialized for generation forward tension on the
elongating axon. Growth cone's cytoskeleton consist of microtubules mostly
located in central domain of the growth cone (lamellipodia),
and actin filaments located in the lamellipodia and finger-like structures
(filopodia). Actin monomers in the peripheral domain undergo constitutive
filament assembly, elongating the lamellipodia and filopodia and pushing the
growth cone membrane in forward direction. Simultaneously actin filaments are
dragged back motors into central domain by myosin-like motors where
the actin filaments depolymerize. The advance of peripheral domain of the
cone determined by the balance of anterograde polymerization and retrograde
retraction of actin. If the balance is shifted toward forward protrusion, the
decrease of in retrograde flow of actin filaments is accompanied by microtubule
polymerization into the peripheral domain, moving the central domain of the
growth cone forward and elongating the axon \cite{Goldberg}.

A great variety of extracellular signals have been found to regulate
axon growth \cite{Nieto}. Extracellular guidance signals can either
attract or repel growth cones, and can operate either at close range
or over a distance.In the literature a family of
chemicals has been found, which in mammals include nerve growth factor (NGF),
brain-derived neurotrophic factor (BDNF), neurotrophins, netrins,
slits, semaphorins, ephrins \cite{Goldberg,Dickson}. Diffusible cues
are netrins, neurotrophins, NGF, BDNF among others. The neuronal growth cone
uses surface receptors to sense these cues and to transduce guidance
information to cellular machinery that mediates growth and turning
responses. These guidance factors we will call axon guidance
molecules (AGM).

Recent studies have shown that electrical activity is required for
growing axons to reach their appropriate target area
\cite{Catalano,Dantzker}. In neurons in culture, the changes in
growth cone motility after electrical stimulation are accompanied by
an influx of calcium through voltage-sensitive calcium channels. The
effects of electrical activity and increases in intracellular
calcium concentration on growth cone morphology are not the same for
all neurons. Some growth cones collapse, some show greater motility
and others do not respond at all, depending on the type of neuron,
the type of neurite (axon versus dendrite) and environmental factors
\cite{Ming}. It has been found that there are several different
signals and signal transduction mechanisms that ultimately result in
alterations of the cytoskeletal structure and growth cone motility.
Modern experimental investigations show that essential role in
controlling growth cone guidance are Ca$^{2+}$ signals. The
Ca$^{2+}$ concentration in growth cones is controlled by various
channels, pumps and buffers. Guidance cues causes the opening of
plasma membrane calcium channels. One of the best studied plasma
membrane channel types on growth cones is voltage-operated calcium
channels. Guidance of axons to their targets probably involves at
least three Ca$^{2+}$ -- dependent effects on motility in particular,
growth promotion, growth inhibition or collapse, and directional
steering (turning). Global Ca$^{2+}$ signals can regulate membrane
dynamics and cytoskeletal elements to control elongation, whereas
localized Ca$^{2+}$ signals can cause asymmetric activation of
downstream effector proteins to steer the growth cone. A small
Ca$^{2+}$ gradient produced by modest Ca$^{2+}$ influx or release
induces repulsion, whereas a larger Ca$^{2+}$ gradient produced by
greater Ca$^{2+}$ influx in combination with release induces
attractive turning \cite{Gomez}. A set of experimental results leads
to "Ca$^{2+}$ set-point" hypothesis: normal growth cone motility
depends on an optimal range of [Ca$^{2+}$] and neurite growth stops
above or below this optimal range \cite{Kater}. Therefore, Ca$^{2+}$
regulation of growth cone motility depends on both the
spatio-temporal patterns of Ca$^{2+}$ signals and the internal state
of the neuron, which is modulated by other signals received by the
neuron. Thus, a rise in [Ca$^{2+}$] in the growth cone activates
numerous target proteins (CaM, CaMKII, myosin, calpain, calcineurin
etc.) and cellular machinery which regulates actin and microtubule
dynamics to provide the growth cone extension and steering
\cite{Henley}.

\subsection{Synaptogenesis}

When a growth cone guided by AGM reaches an appropriate target cell,
the synaptogenesis and synapse refinement processes start. Synapse
formation is controlled by dynamic interactions between various
genes and their encoded proteins and occurs throughout development
to generate synapse specificity \cite{Munno,Garner}. The modern
version of Dale's principle suggests that all of a particular
neuron's terminals release the same set of neurotransmitters
\cite{Eccles}.
 Nevertheless, it is well known that a neuron can store and
presumably release the different sets of transmitters from
individual axon endings \cite{Sossin}. The neurotransmitter choice
of the neurons depends on programmed and environmental factors, and
this process is neither limited by a critical period nor restricted
by their insertion in a network \cite{Gomez-Lira}. Calcium transient
patterns plays a key role in the differentiation of neural precursor
cells, and their frequency may specify neuronal morphology and
acquisition of neurotransmitter phenotype \cite{Ciccolini}. Neuronal
activity also plays a main role in the neuronal connection
establishment \cite{Andras}- \cite{Marom} %\cite{AndrasGoulding,Marom}
and refinement
\cite{Kandler} processes. The electrical activity of neurons can
regulate the choice of neurotransmitter in cultured neurons through
calcium influx and can differentially affect the regulation of
transmitter expression \cite{Gutierrez}. Certain neurons choose the
neurotransmitter which they use in an activity-dependent manner, and
different trophic factors are involved in this phenotype
differentiation during development. Regulation of transmitter
expression occurs in a homeostatic manner. Suppression of activity
leads to an increased number of neurons expressing excitatory
transmitters and a decrease number of neurons expressing inhibitory
transmitters and vice verse \cite{Borodinsky}. Based on the above
discussion we assume that each neuron's  axon can release
different neurotransmitters and can establish different types of
synaptic connections (inhibitory or excitatory). The type of synapse
can be determined by state of presynaptic or/and postsynaptic
neuron. For simplification we assumed that the type of a synaptic
connection between cells depends on the state of postsynaptic cell
at synaptogenesis process.

\subsection{Activity dependent AGM release}
Release of some neurotrophic factors  can be triggered by external
stimulation and neuron's electrical activity
\cite{Hartmann,Balkoweic}. The activity dependent release of AGM's
is a key assumption in our model. We doubt there is  complete proof of activity
dependent AGM's
release and we consider this point a an hypothesis.

\section{The model}

Let us proceed to description of the model adopted here. The concentration of
AGM, $c_i = c_i(\textbf{r} - \textbf{r}_i,t)$, at point $\textbf{r}$ in the
moment
$t$ released by the $i$-th neuron at point $\textbf{r}_i$ can be found as the
solution of the equation
\begin{equation}
 \frac{\partial c_{i}}{\partial t} - D^2 \triangle c_{i} + k c_{i} =
J_i(\textbf{r},t).
\end{equation}
Here, $D^2$ and $k$ are AGM diffusion and degradation coefficients in the
intracellular medium respectively. The source $J_i$ is amount of the AGM per
unit time. It is well-known that the solution of this equation has the following
form
\begin{equation}
 c_i(\textbf{r} - \textbf{r}_i,t)= \int G_d (\textbf{r} - \textbf{r}',t)
c_i^0(\textbf{r}')d\textbf{r}' + \int_0^t dt'\int G_d (\textbf{r} -
\textbf{r}',t-t') J_i(\textbf{r}',t')d\textbf{r}',
\end{equation}
where $c_i^0(\textbf{r})$ is initial distribution of the concentration, $d$ is
dimension of the problem and the the Green function has standard Gaussian form
\begin{equation}
 G_d(\textbf{r},t) = \frac{1}{(4\pi t D^2)^{d/2}}
e^{-kt -\frac{\textbf{r}^2}{4tD^2}}.
\end{equation}
We suppose that in the initial time there is no AGM and the process is managed
by the source
\begin{equation}
 J_i(\textbf{r},t) = a \delta^{(d)} (\textbf{r} - \textbf{r}_i) j_i(t),
\end{equation}
which is concentrated at the $i$-th neuron. The parameter $a$ describes
amount of AGM released per unit second. This quantity means the amount of
AGM per unit time. The $j_i(t)$ describes activity of the neuron and $j_i(t)
\leq 1$.

Using this information we have the following form of the concentration
\begin{equation}
 c_i(\textbf{r} - \textbf{r}_i,t)= a\int_0^t dt' G_d (\textbf{r} -
\textbf{r}_i,t-t') j_i(t').\label{eqconc}
\end{equation}

For description of neural electrical activity several models
have been developed \cite{Rabinovich}.  For simplicity we take the
activity $j_i(t)$ is the subject for the equation \cite{Voges}
\begin{equation}
 \tau \frac{dj_i(t)}{dt} = -j_i(t) + f\left(j^{ext}_i(t) + \sum_{k=1,k\not =i}^N
\omega_{ik} j_k(t)\right),\label{eqactiv}
\end{equation}
where the functional $f$ has step form
\begin{equation}
 f(x) = x \theta (x),
\end{equation}
with step function $\theta$ and $N$ is amount of neurons.
The matrix $\Omega$ with elements
$\omega_{ik}$ describes the influence the $k$-th neuron to $i$-th neuron;
$\omega_{ik} = 1$ means excitatory and $\omega_{ik} = -1$ inhibitory
connections, for $\omega_{ik} = 0$ there is no influence.  The functions
$j_i^{ext}(t)$ describe the external sources which excites $i$-th neuron. Now we
define the vector $\textbf{g}_i(t)$ which describes the tip of the axon which
started to grow from $i$-th neuron. It is subject for equation
\begin{equation}
 \frac{d\textbf{g}_i}{dt} = \lambda {\cal F}(j_i)\sum_{k=1}^N \nabla
c_k(\textbf{g}_i - \textbf{r}_k,t),\label{eqaxon}
\end{equation}
where functional ${\cal F}$ has the form of the step function, too:
\begin{equation}\label{threshold}
 {\cal F}(j) = \theta (j^{th} - j).
\end{equation}
The functional $\cal{F}$, in fact, is smooth function of activity $j$. In our
model we adopt the simplest form of this functional as a step-function by using
the threshold parameter. It means that the axon is quiescent if the activity of
its neuron is greater then some threshold value $j^{th}$. The parameter
$\lambda$  is a coefficient describing axon's sensitivity and motility.

At the initial moment we set matrix $\Omega = 0$ which means no connections
between neurons. Then we solve the above equations (\ref{eqaxon}),
(\ref{eqactiv}), (\ref{eqconc}) assuming that some neurons are exited by
external force that is assuming that some of $j^{ext}_i(t)$ are not zero. We
obtain position $\textbf{g}_i$ of the axon's tip at the moment $t$. If
$i$-th axon makes connection with $k$-th neuron we set $\omega_{ik} = \pm 1$.

\section{Results}
In this section we present the result of the numerical simulation of
the model considered above. We simulate two ($N=9$ neurons) and three
($N=27$ neurons) dimensional realizations of the model.  For simplicity we
consider the network of neurons in form of the lattice with increment
(distance between neurons) $d = 0.5\ mm$. Each neuron has single
growth cone, which we consider as its axon. Initially all growth
cones are located near its soma, and all synaptic weights are equal
to zero ($w_{ik}=0$ $i,j=1...N$), which means no connections between neurons.
The system of differential equations
(\ref{eqaxon}), (\ref{eqactiv}), (\ref{eqconc}) were
integrated simultaneously by the Euler method. The parameters
used in the model were taken from different experimental findings
and we list them below.
\begin{enumerate}
 \item The AGM diffusion coefficient $D^2 = 6\cdot 10^{-7}cm^2/s$, \cite{Goodhill1997}.
 \item The amount of AGM per unit second $a = 10^{-5} nM/s$,
\cite{Gundersen,Goodhill1997}.
 \item The relaxation time of activity $\tau = 10\ ms$, \cite{Voges}.
 \item The coefficient describing axon's sensitivity and motility $\lambda
=4\cdot 10^{-6} cm^2/nM\cdot s$, \cite{Rosoff}.
\end{enumerate}

The threshold parameter $j^{th}$ is a parameter exclusive to our model and for
this reason there is no experimental value that can be assumed to it. We set the
threshold parameter, $j^{th} = 0.51$, to take good agreement with
experimental observation of the network growth \cite{Segev2003}. The
degradation coefficient $k$ may be found, in principal, from
specific experiments. Unfortunately, there is no information about
this parameter and we put $k = 10^{-3} s^{-1}$ to be in agreement
with network growth. This parameter regulates the rate of
the axon's growth. The greater $k$ the smaller rate of axon's
growth.

There is another threshold parameter of activity which defines sort of
connections: $\omega = +1$ or $\omega = -1$. For homeostasis we take this
parameter to equal $0.51$. If activity of postsynaptic neuron's greater then $0.51$ we set
$\omega = -1$ and vice versa. This is a crude approximation but it is
enough to describe the neurons networks development.

\begin{figure}[ht]
\begin{center}
{\hspace*{-6ex}\epsfxsize=8truecm\epsfbox{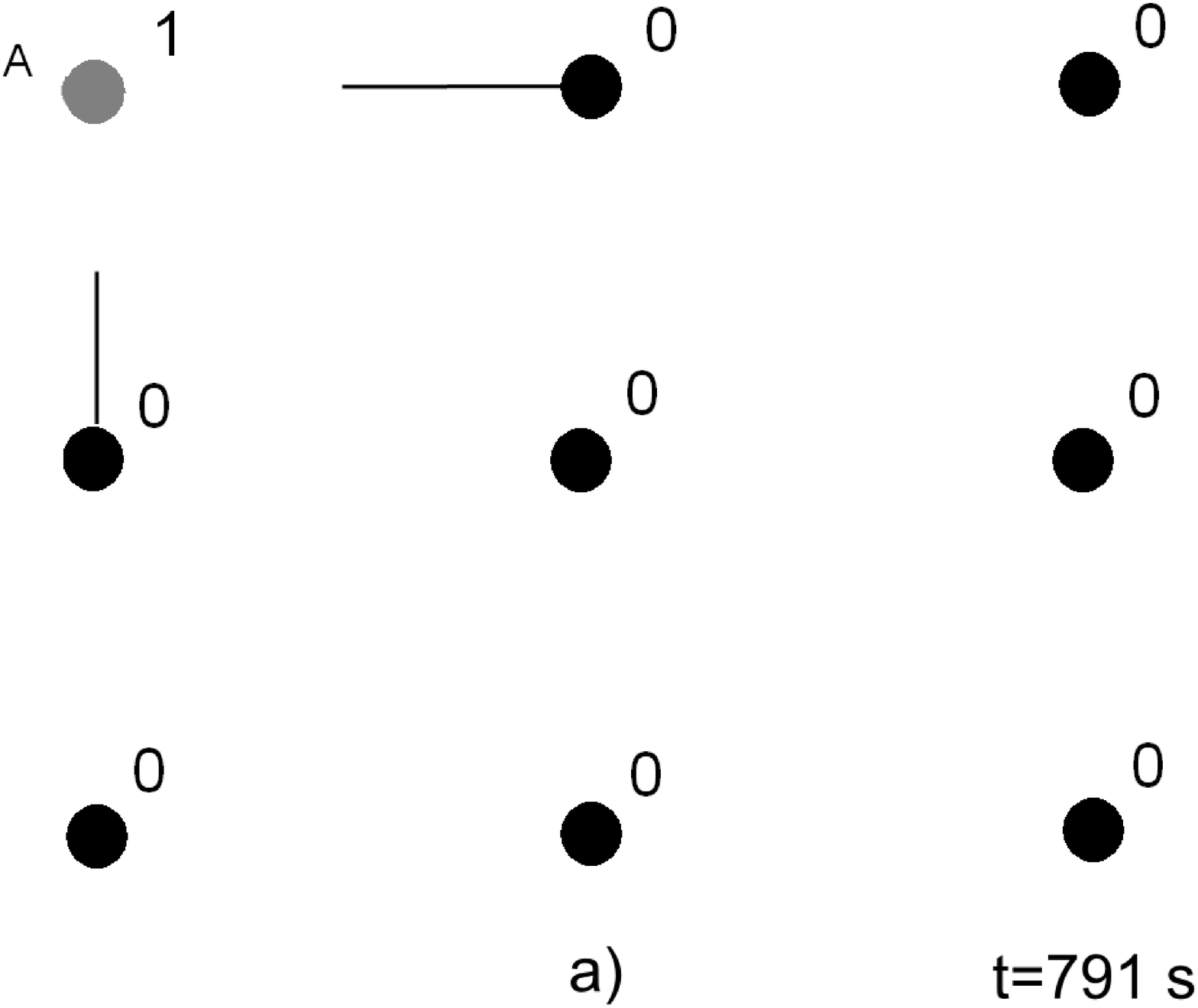}}%
{\hspace*{-3ex}\epsfxsize=8truecm\epsfbox{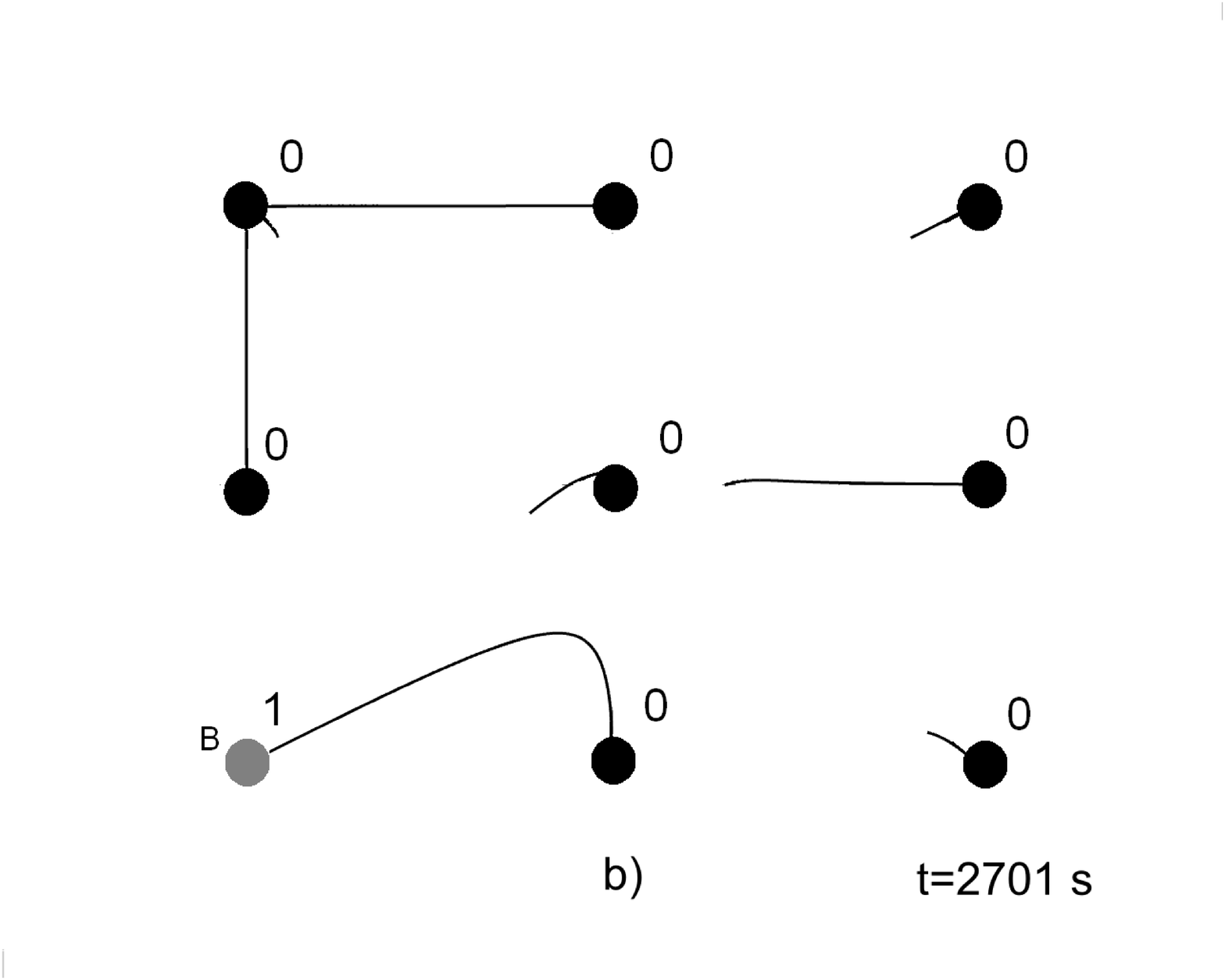}}

{\hspace*{-6ex}\epsfxsize=8truecm\epsfbox{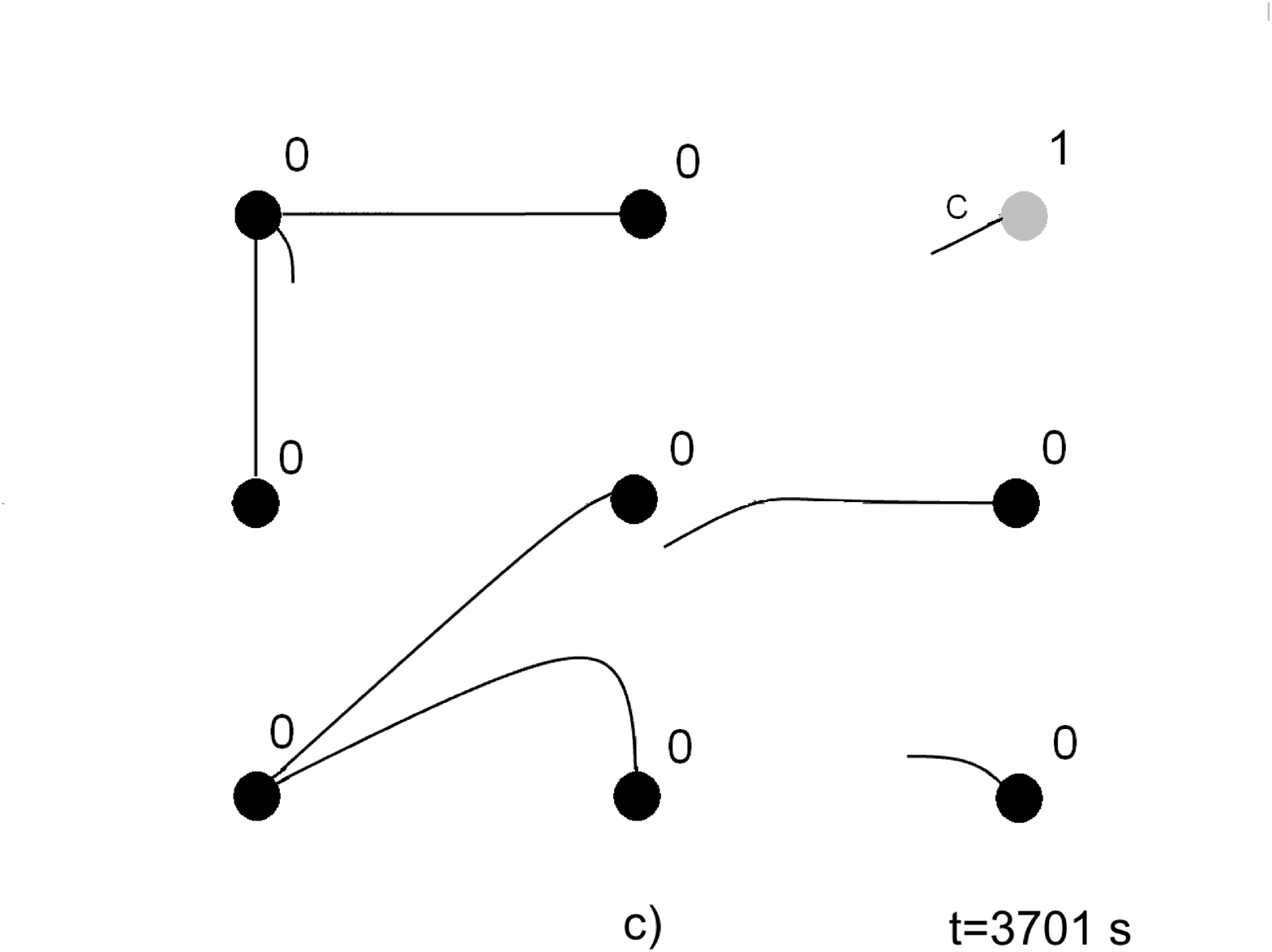}}%
{\hspace*{-3ex}\epsfxsize=8truecm\epsfbox{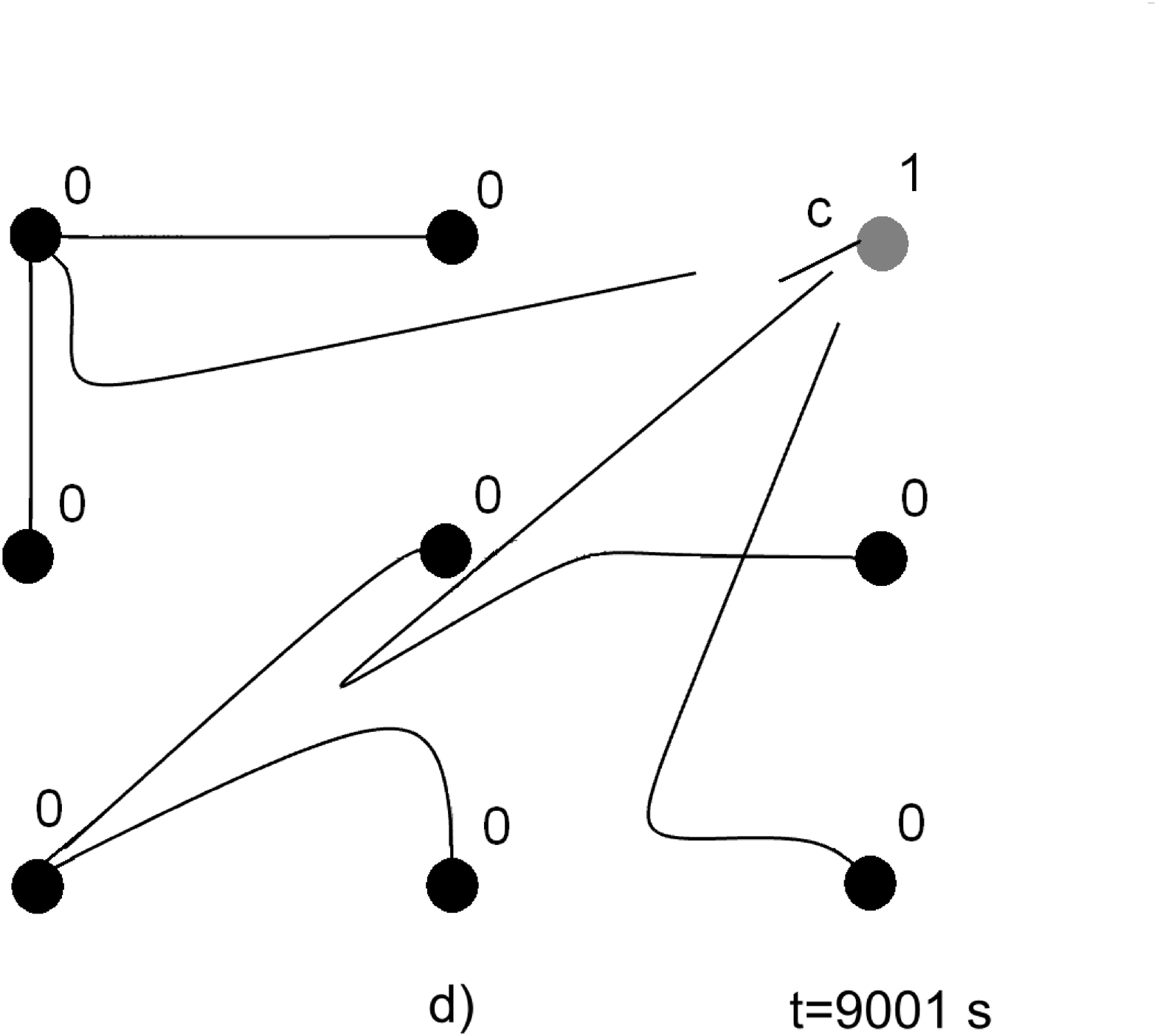}}
\end{center}\caption{The numerical simulation the development of
the two-dimensional neuron's network. The neurons are shown as circles. The
blackness of the circles proportional to activity: zero activity plotted by
black and maximal activity is grey. The number close to neurons means activity
of them. The black lines are axons. We show snapshots of neuron
network for different moments of time namely,
$t=791s,2701s,3701s,9001s$.}\label{D2}
\end{figure}

As an example we present in Fig. \ref{D2} and \ref{D3} the neural network which
was obtained by using the following training pattern sequence. (We show
snapshots of the dynamic process; full animation may be found in URL:
http://neurowiring.narod.ru/video.html)

\subsection{2 dimensional case}
Neuron A was being active with activity $1$ during the time period from $0s$ to
$800 s$ and activity of other neurons in this period is zero. In
Fig. \ref{D2}a we show the snapshot of the system at the moment
$t=791s$. We observe that the nearest two neurons growth cones grow
to this active neuron A. After some time the growth cones reached
the active neuron A and synaptical connects appear $w_{1,2}=1,
w_{1,4}=1$. The growth cones of the distant neurons do not appear
because concentration of the AGM far from the active neuron is
neglegible.

In the interval $t=800s - 1200s$ the central neuron was active with activity
$1$. The neuron B was active in interval $t=1200s - 3500s$. In Fig.
\ref{D2}b we reproduce the snapshot of the system at the moment $2701s$. We
note that the growth cones of neurons close to the another neuron did not
reach this central neuron and the neuron B became active. The consequence of
this fact is curved form of the axons.

Since the moment $t=3500s$ the neuron C plays the role due to its
activity (see Fig. \ref{D2}c and \ref{D2}d). We show the snapshots
of the system at the moments $3501s$ and $9001s$. We observe the the
new connections to neuron B appeared $w_{3,5}=-1, w_{3,6}=-1$. The
growth cones which did not reach neuron B turned to the active
neuron C. The growth cone of the active neuron C which started to
grow in the earliest moment is quiescent in the period of its
activity (see Fig. \ref{D2}c and \ref{D2}d). Mathematically it is
described by threshold function $\cal F$ (\ref{threshold}).

We observe that the topology of the neural network strongly depends on the sequence
of neuronal activity the reason of which is external influence. In real system
the growth of the connections in the complex connection structure will depend
on the internal oscillation of activity rather then external signals. The
external influence is very important at the initial time during the network
development.

\section{3 dimensional case}

\begin{figure}[ht]
\begin{center}
{\epsfxsize=8truecm\epsfbox{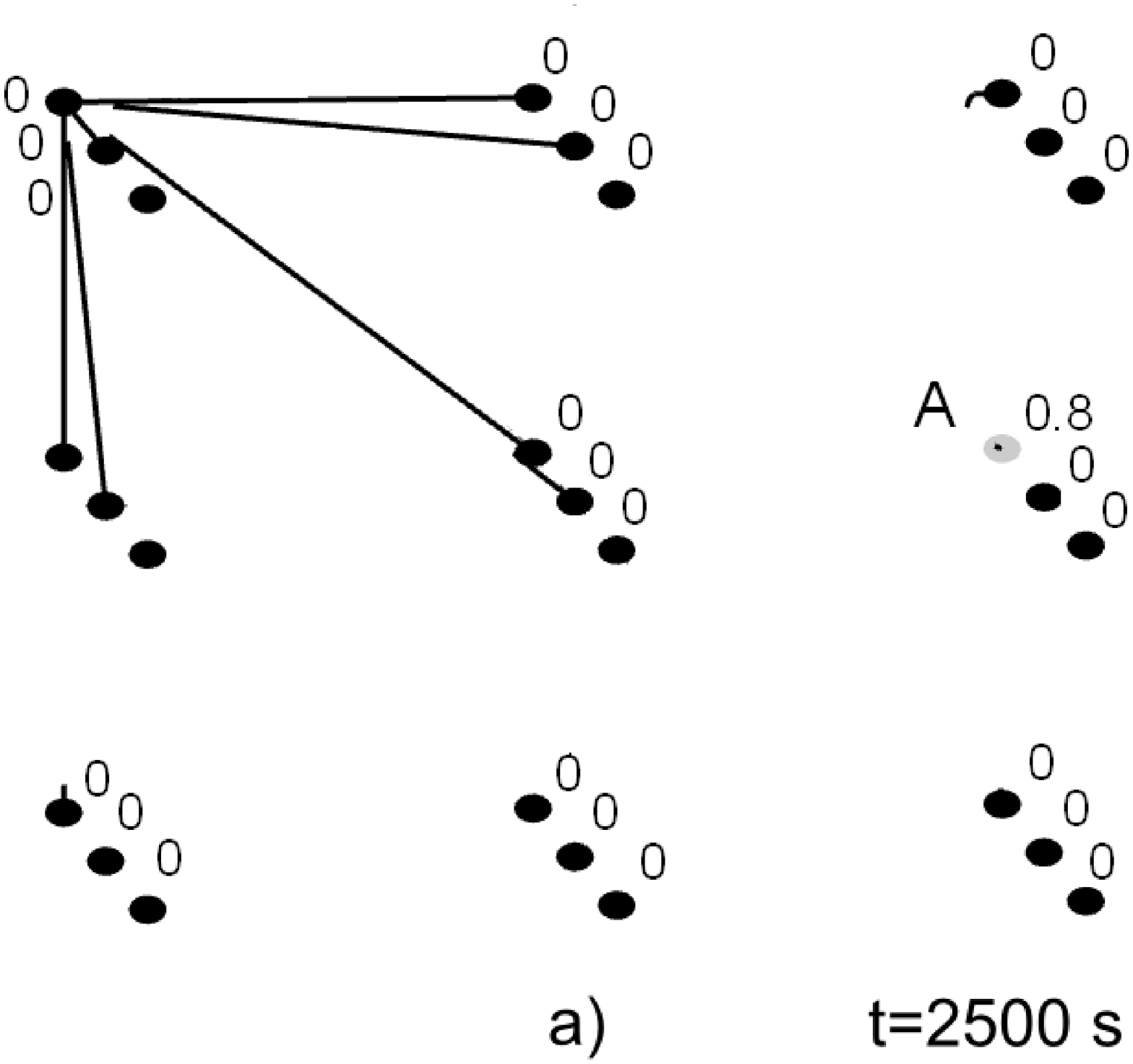}}%
\hspace*{-4ex}{\epsfxsize=8truecm\epsfbox{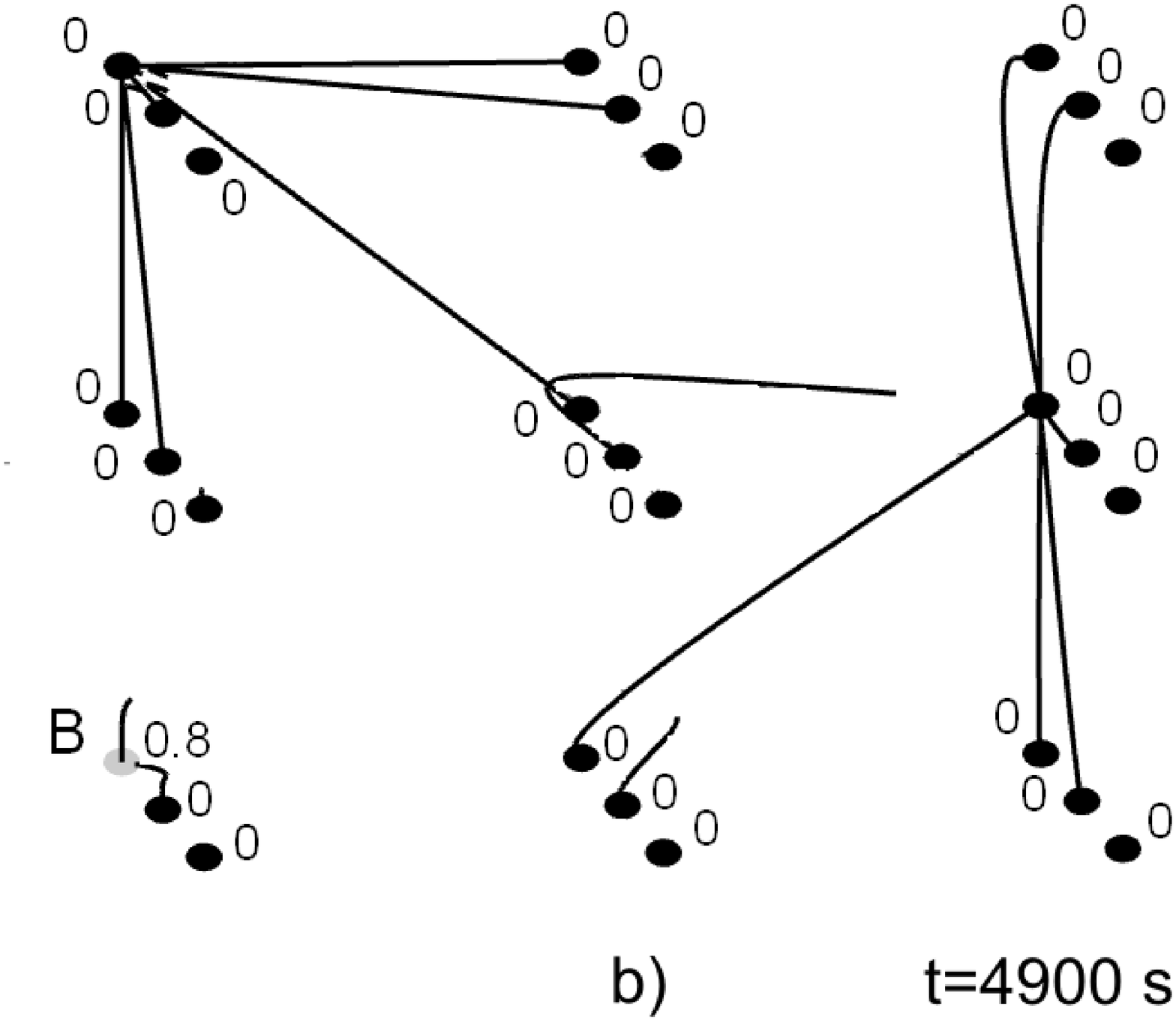}}

{\epsfxsize=8truecm\epsfbox{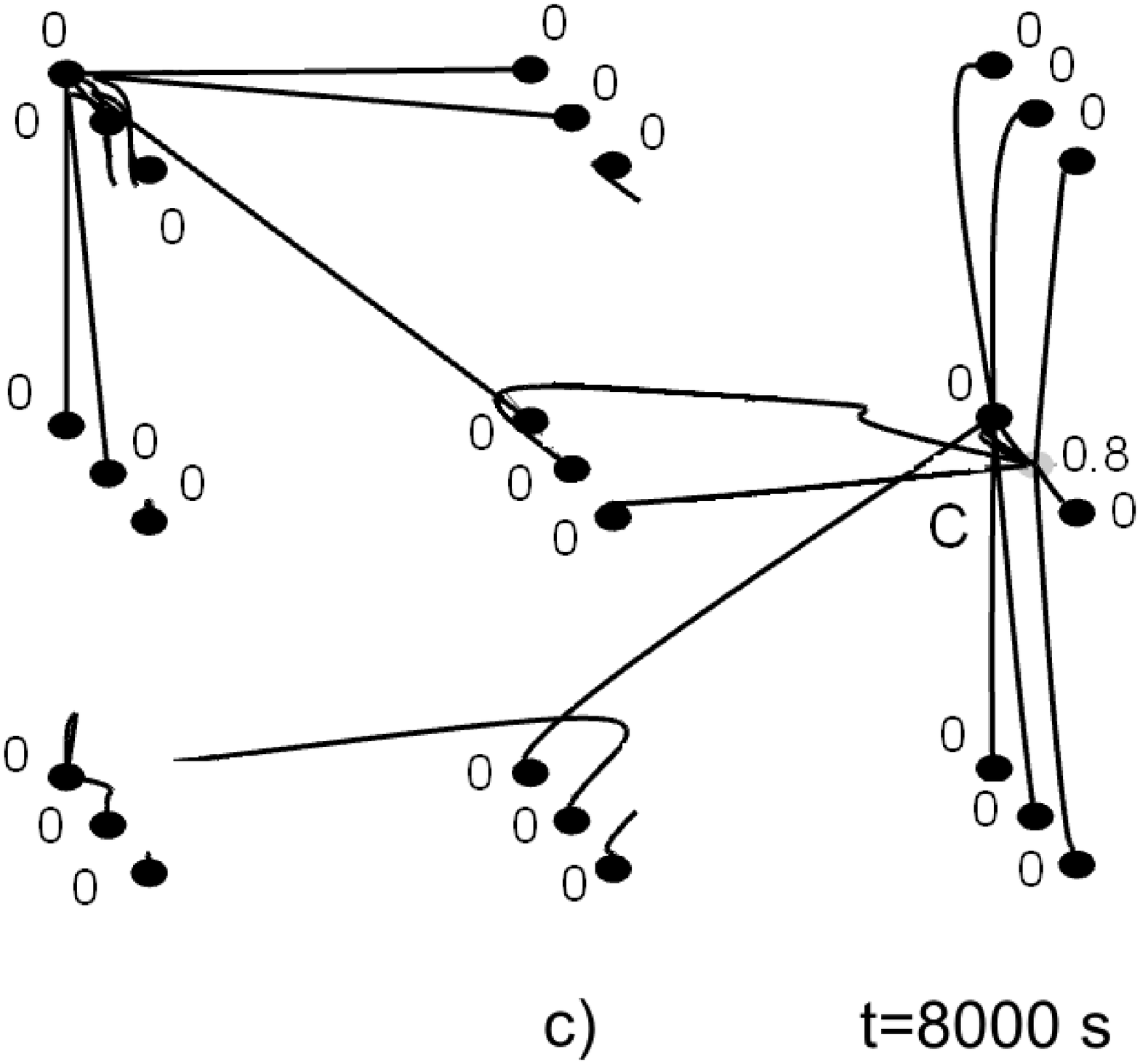}}%
\hspace*{-4ex}{\epsfxsize=8truecm\epsfbox{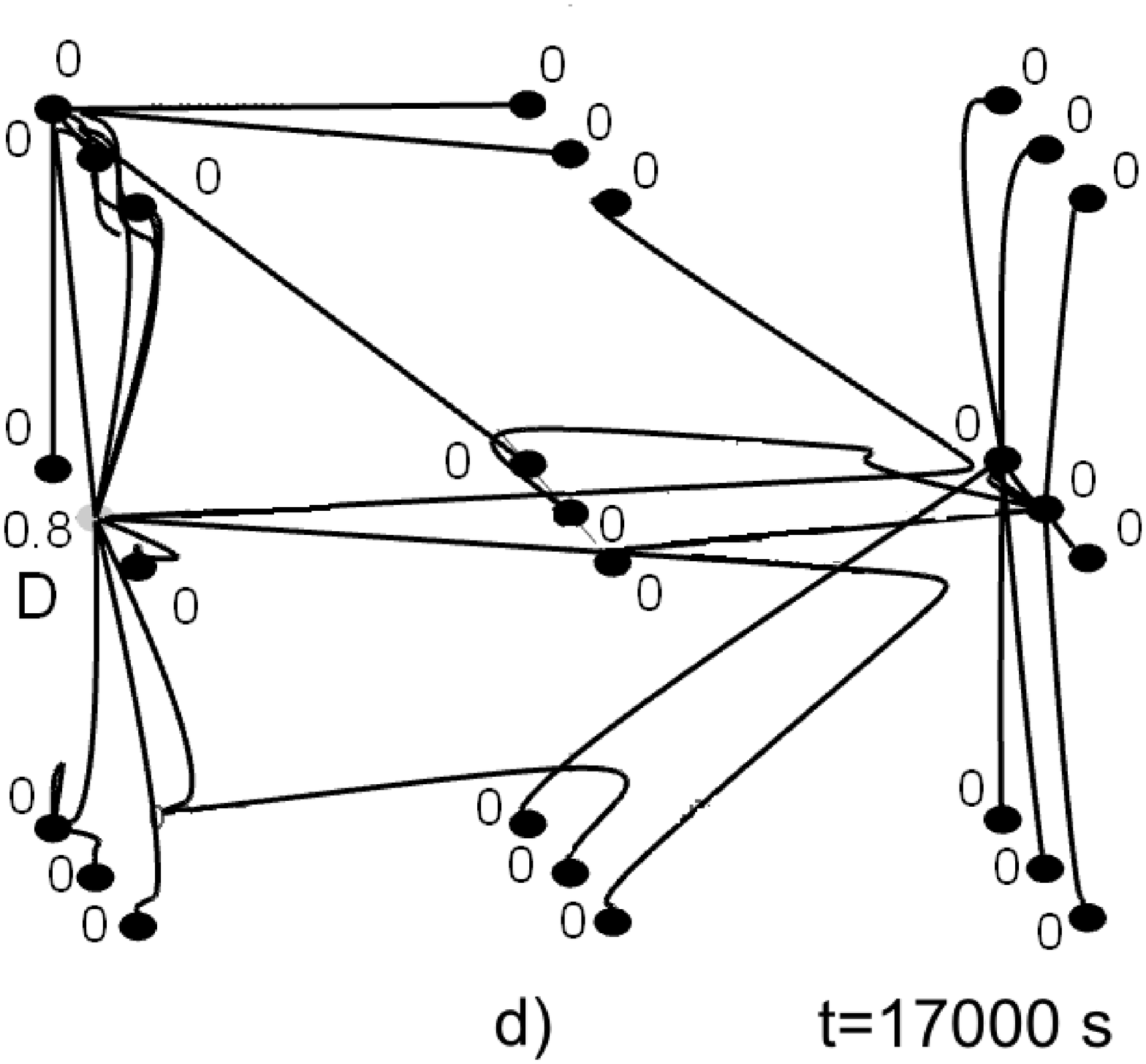}}
\end{center}\caption{The numerical simulation the development of the
three-dimensional neuron's network. The neurons are shown as balls. The
blackness of the balls proportional to their activity: zero activity plotted by
black and maximal activity is grey. The number close to neurons means activity
of them. The black lines are axons. We show snapshots of neuron
network for different moments of time namely,
$t=2500s,4900s,8000s,17000s$.}\label{D3}
\end{figure}

The situation in this case is close to the 2 dimensional case. The first neuron in
the upper cone was active in the period $t=0s - 2000s$ with activity $0.8$. In
the period $t=2000s - 4000s$ the neuron A was active. The snapshot for moment
$2500s$ is shown in Fig. \ref{D3}a. In the period $t=4000s - 6000s$ the
neuron B was active (see Fig. \ref{D3}b for $t=4900s$). The neuron C was active
in the period $t=6000s - 9000s$ (see Fig. \ref{D3}c for $t=8000s$). The neuron
D starts to be active since moment $9000s$ (see Fig. \ref{D3}d for $t=17000s$).

We observe the network development close to 2 dimensional case. The topology of
the network depends on the sequence of neuronal activity.

\section{Discusion}

In above sections we developed the new theoretical approach to
describe the growth of the network of neurons. The model is based on
the diffusion of the AGM and dynamic of the diffusion satisfies to
diffusion equation. The rate of the growth cone movement is
proportional to concentration gradient which should be the case. The
peculiarity of the model is that the activity of the neuron manages
the release of the AGM which influence to axon's growth. This
process leads to appearance of the new connections in the system and
network development. In framework of the model and with real
parameters obtained in experiments we have correct picture of
network growth (see Fig. \ref{D2} and \ref{D3}) in temporal scale.
In particular, the  Fig. 2 ($t=17000 s$), shows some neurons to have
only long-range connections without any local connections. This
showes how nested neural networks are formed in the cerebral cortex.

Our expectations of the network topology and direction of the axon's
growth were realized in framework of the model. We confine ourself
for two reasons. First of all we would like to consider in detail
the dynamics of the axon's growth in dependence of the neurons
activity. Second, the real system contains huge amount the neurons.
The present calculations are confined by power of our computers. In
the future we intend to make numerical simulations for more real
amount of the neurons and axons.

Real cortical networks have more complex structure comparing with that obtained
in framework of our model. The neural network development is controlled by many
factors which are out of scope of our model, namely, cell adhesion molecules,
multiple guidance factors and etc. In the model presented here we considered
only one of them, the axon guidance by single diffusable factor, which is, in
fact, the most important factor in activity dependent development.  In our model
axons have only one branch and single growth cone, without taking into account
axon branching. The cortical neuron's axons have plenty branched structure and
different branches of single axon grow to different target neurons. In the model
each neuron can make connection only with one target neuron. Alongside with
guidance growth cone by chemoattraction the process is controlled also by
chemorepellants. Different types of neurons have different properties and
cortex consists of a lot amount of different neurons. In the model
presented here we take into account only one type of the neurons. The guidance
of particular growth cones to the target is not controlled by single AGM. In
particular parts of growth cone trajectory the different AGM's take part in
axon guidance. The growth cones can go to long distances and
they result to long-range intra-cortical and cortico-cortical connections.

Real neurons possess also the strong branched dendrite structure. The dendrite
growth represents very complex process which is managed by many factors. In
our model the axons' growth cones make synaptic connections directly to a soma.
The most of synaptic connections in cortex are on dendrites which are absent in
our model. To obtain a more precise picture of cortex development we should
take into account also the morphological propertied of neurons. We plan to
include dendrites and morphological propertied of neurons into our model in
future investigations.
\section{Conclusions}

The theoretical framework developed here can be used for description
the development of a particular set of neurons constituting  the
neural system. In this paper we show that chemotactically guidance
of growth cones by AGM released by neurons can be a basis of
neural networks growth, topology and development. The concentration
of AGM released by individual cells is basis for correct axon
guidance with appropriate rate. All parameters describing the model
are taken from different real experiments. This model can be used
for description the growth of developing neural network. It is well
known that the wiring of neural networks takes place on the
chemotaxis based axon guidance. The connections structure between
neurons is very complex in real networks. Such complex structure can
appear only by switching of and switching of the chemical signal
that regulates the growth of axon. Using this model we conclude that
these process can be based for learning, because creation of new
connection leads to increasing of network complexity in structure.
Another application of this model is in treatment of damaged neural
tissue by stem cells. Using this model we may describe the processes
of integration of stem cells into existing network. This model can
be used also for understanding the processes which takes place at
deep brain stimulation by electrical current. We showed that the
electrical stimulation of individual cells leads to alternation of
AGM release and in growth cones guidance, and deep brain stimulation
can change the network structure. Another application of our model
can be in modeling of imprinting memories in cultured neural
networks  \cite{Baruch}.

\end{document}